\documentclass[aip,reprint] {revtex4-1}

\usepackage{upgreek}
\usepackage{amsmath}
\usepackage{amsfonts}
\usepackage{amstext}
\usepackage{mathtools}
\usepackage{amssymb}
\usepackage{siunitx}
\usepackage{color}
\usepackage[normalem]{ulem}

\newcommand{\vx}{{\boldsymbol{x}}}
\newcommand{\vp}{{\boldsymbol{p}}}
\newcommand{\vq}{{\boldsymbol{q}}}

\newcommand*{\mean}[1]{\left< #1 \right>}
\newcommand*{\e}{\mathrm e}

\newcommand{\CV}{x}  
\newcommand{\CC}{s}  
\newcommand{\euler}{\mathrm{e}}
\newcommand{\kb}{k_\mathrm{B}}
\renewcommand{\d}{\mathrm{d}}
\renewcommand{\vec}[1]{{\boldsymbol{#1}}}

\newcommand{\itGamma}{{\it{\Gamma}}}	
\DeclareSymbolFont{uplargesymbols}{OMX}{mdbch}{m}{n}
\SetSymbolFont{uplargesymbols}{bold}{OMX}{mdbch}{b}{n}
\DeclareMathSymbol{\upintop}{\mathop}{uplargesymbols}{82}
\makeatletter\newcommand{\upint}{\DOTSI\upintop\ilimits@}\makeatother
\usepackage{lipsum}
\definecolor{gray}{gray}{0.5}

\usepackage{stmaryrd} 

\definecolor{newblue}{rgb}{0.5,0.2,1}

\usepackage[normalem]{ulem}
\newcommand{\stkout}[1]{\ifmmode\text{\sout{\ensuremath{#1}}}\else\sout{#1}\fi}

\newcommand{\SIPCAEQ}		{S1}
\newcommand{\SIFELsV}    	{S2}
\newcommand{\SIConv}       	{S3}
\newcommand{\SIFELsVneq}    	{S4}
\newcommand{\SIFELneqPC}        {S5}
\newcommand{\SIDiheNEQ}         {S6}

\begin{document}
\author{Matthias Post, Steffen Wolf and 
  Gerhard Stock}
\email{stock@physik.uni-freiburg.de}
\affiliation{Biomolecular Dynamics, Institute of Physics, Albert Ludwigs
University, 79104 Freiburg, Germany}
\title{Principal component analysis of nonequilibrium molecular
  dynamics simulations} 
\date{\today}

\begin{abstract}
  Principal component analysis (PCA) represents a standard approach to
  identify collective variables $\{x_i\}\!=\!\vx$, which can be used
  to construct the free energy landscape $\Delta G(\vx)$ of a molecular
  system.  While PCA is routinely applied to equilibrium molecular
  dynamics (MD) simulations, it is less obvious how to extend the
  approach to nonequilibrium simulation techniques. This includes,
  e.g., the definition of the statistical averages employed in PCA, as
  well as the relation between the equilibrium free energy landscape
  $\Delta G(\vx)$ and energy landscapes $\Delta{\cal G} (\vx)$
  obtained from nonequilibrium MD. As an example for a nonequilibrium
  method, ``targeted MD'' is considered which employs a moving
  distance constraint to enforce rare transitions along some biasing
  coordinate $s$. The introduced bias can be described by a weighting
  function $P(s)$, which provides a direct relation between
  equilibrium and nonequilibrium data, and thus establishes a
  well-defined way to perform PCA on nonequilibrium data. While the
  resulting distribution ${\cal P}(\vx)$ and energy
  $\Delta{\cal G} \propto \ln {\cal P}$ will not reflect the
  equilibrium state of the system, the nonequilibrium energy landscape
  $\Delta{\cal G} (\vx)$ may directly reveal the molecular reaction
  mechanism. Applied to targeted MD simulations of the unfolding of
  decaalanine, for example, a PCA performed on backbone dihedral angles
  is shown to discriminate several unfolding pathways. Although the
  formulation is in principle exact, its practical use depends
  critically on the choice of the biasing coordinate $s$, which should
  account for a naturally occurring motion between two well-defined
  end-states of the system.
\end{abstract}
\maketitle

%
%

\section{Introduction}

The calculation of the free energy landscape of a molecular system
along some reaction coordinate $\vx$ represents a central task of {\em
  in silico} modeling. Employing unbiased molecular dynamics (MD)
simulations, the free energy landscape $\Delta G(\vx)$ can be directly
calculated from the probability distribution $P(\vx)$ via
\begin{align} \label{eq:pmf} 
  \beta \Delta G(\vx) &= -  \ln \left[P(\vx)/ P(\vx_0) \right] ,
\end{align}
where $\beta \!=\! 1/k_{\rm B} T$ is the inverse temperature and
$\vx_0$ refers to some reference state. Given a suitable choice of
$\vx$, the free energy landscape reveals the relevant regions of low
energy (corresponding to metastable states) as well as the barriers
(accounting for transition states) between these regions, and may
therefore visualize the pathways of a biomolecular
process.\cite{Onuchic97, Dill97,Wales03} To
identify optimal reaction coordinates $\vx \!=\! (x_1,\ldots,x_{d})$,
often referred to as collective variables $x_i$, various
dimensionality reduction methods have been
developed,\cite{Rohrdanz13,Peters16,Noe17,Sittel18} a popular example
being principal component analysis (PCA).\cite{Amadei93,Mu05}

Standard unbiased MD simulations become impractical, if the states are
separated by high energy barriers such that transitions between them
occur only rarely. To this end, a number of enhanced sampling
techniques\cite{Chipot07,Christ10,Fiorin13,Tribello14,
  Sugita99,Grubmueller95,Rico19,Laio02,
  Darve08,Torrie77,Isralewitz01,Park04} have been proposed, including,
e.g., replica-exchange MD, \cite{Sugita99} conformational
flooding,\cite{Grubmueller95} metadynamics,\cite{Laio02} and adaptive
biasing force sampling.\cite{Darve08} To enforce rare transitions, in
particular, one may employ some external force to pull the molecule
along some --usually one-dimensional-- coordinate $s$.  Various
versions of this nonequilibrium technique exist, including simulations
using moving harmonic restraints \cite{Torrie77} along
$s$ such as steered MD\cite{Isralewitz01,Park04} or constrained
simulations \cite{Sprik98,Ciccotti05} such as targeted MD (TMD)
simulations,\cite{Schlitter93A,Schlitter94,Schlitter01} which employ
moving distance constraints. While our study in principle applies to
all these methods, to be specific we here focus on TMD.\cite{note4}

From these externally driven nonequilibrium simulations, the free
energy profile $\Delta G(s)$ can be calculated in various ways. In the
quasi-static limit of very slow pulling, we may perform equilibrium
calculations of the free energy for selected values of $s$. This is
the basis of thermodynamic integration, which calculates the free
energy difference $\Delta G(s)= G(s) - G(s_0) $ via the potential of
mean force, \cite{Berendsen07}
\begin{align}\label{eq:TI}
\Delta G(s) = \upint_{s_0}^{s} \!\! \d s^\prime \, \frac{\mathrm{d}
	G}{\mathrm{d} s'} 
=  \upint_{s_0}^{s} \!\! \d s^\prime \left \langle {f}_\text{c}(s') \right
\rangle_{{\rm eq}}
\end{align}
where $\left \langle {f}_\text{c} (s) \right \rangle_{{\rm eq}}$
represents an equilibrium average of the pulling force ${f}_\text{c}$
at point $s$. While representing a straightforward and well-established
approach to compute $\Delta G$, thermodynamic integration is
in practice quite demanding, because it typically requires
numerous and relatively long MD simulations
to converge to equilibrium.

Alternatively we may calculate the free energy directly
from the nonequilibrium pulling trajectories by employing Jarzynski's
equality\cite{Jarzynski97}
\begin{align}\label{eq:Jarzynski}
 e^{-\beta \Delta G(s)}  = \left\langle e^{-\beta W(s)} 
                                \right\rangle_{\rm neq}.
\end{align}
Here $ \langle \cdot \rangle_{\rm neq}$ denotes an ensemble average
over {independent realizations of the pulling process starting from an
  equilibrium distribution at $s=s_0$, and
\begin{align}\label{eq:Ws}
W(s) =  \upint_{s_0}^{s} \!\! \d s' \, f_{\rm c}(s') 
\end{align}
represents the work performed on the system by external pulling.
Since the pulling coordinate $s$ represents the control parameter of a
{\em constrained} simulation, for TMD Jarzynski's identity directly yields the
free energy profile.\cite{Mulders96} Note that this equivalence does
not hold for {\em restrained} simulations, where the system is allowed
to fluctuate around the value of $s$ and the free energy has to be
recovered by other means.\cite{Kumar92,Hummer01a,Hummer05a}
Various ways to compute the exponential average in Jarzynski's
identity have been suggested,\cite{Hendrix01,Park04,Oberhofer09,
Dellago14,Wolf18} including a ``fast growth''
implementation\cite{Hendrix01} and a cumulant expansion
\cite{Hendrix01,Park04,Wolf18} of Eq.\ \eqref{eq:Jarzynski},
\begin{align}\label{eq:dGcum}
\!\! \Delta G (s) = \left\langle W(s) \right\rangle_{\rm neq}  -
  \frac{\beta}{2} \left\langle (W(s)\!-\!\left\langle W(s)
  \right\rangle_{\rm neq})^2 \right\rangle_{\!\rm neq} ,
\end{align}
which approximates the dissipated energy by the variance of the work. 

Given an optimal choice of the pulling coordinate that is similar to
the motion in the unbiased process, the one-dimensional free energy
profile $\Delta G(s)$ may already describe the biomolecular reaction
correctly and in desired detail. By constraining only a single
coordinate, however, the system is free to move in the remaining
degrees of freedom and may, e.g., sample important intermediate
states. For example, when we pull a ligand out of a protein binding
pocket, several unbinding pathways may occur, whose description
requires additional coordinates. Similar as in the case of unbiased MD
simulations, it is therefore desirable to employ some dimensionality
reduction approach such as PCA, in order to describe the energy
landscape along an appropriate reaction coordinate $\vx$. While PCA
is routinely applied to unbiased equilibrium MD simulations, the
situation is less obvious for biased nonequilibrium techniques such as TMD. This includes, e.g., the definition of the
statistical averages employed in the PCA, as well as the relation of
the free energy landscape obtained from equilibrium simulations and
energy landscapes obtained from nonequilibrium MD.

In this work we consider the calculation of multidimensional energy
landscapes from TMD simulations. In particular, we demonstrate the
application and interpretation of PCA of nonequilibrium data. Adopting
decaalanine in vacuo as a well-established model problem to test
TMD,\cite{Park03,Procacci06,Forney08,Oberhofer09,Hazel14} we compare
and analyze unbiased MD and TMD data.

%
%
\section{Theory and methods}
\subsection{Free energy landscapes from constrained dynamics}

In general, the probability distribution of a variable is obtained by
inserting a $\delta$-function into the partition function. In the case
of unbiased MD simulations in the canonical ensemble, for example, the
probability distribution of reaction coordinate $\vx$ used in Eq.\
(\ref{eq:pmf}) is given by
\begin{align} \label{eq:Px} 
P(\vx) &= Q_{\rm eq}^{-1} \upint \!\! \d \vq \d \vp \; \e^{-\beta H (\vq,\vp)}
         \, \delta(\vx\!-\!\vx(\vq)) \nonumber \\
  &\equiv \left\langle \delta(\vx\!-\!\vx(\vq)) \right\rangle_{\rm eq},
\end{align}
where $(\vq,\vp)$ denote the phase-space coordinates of the system's
microstate, $H$ represents its Hamiltonian, and
$Q_{\rm eq}=\upint \!\! \d \vq \d \vp \; \e^{-\beta H (\vq,\vp)}$ its
partition function.

In the case of TMD simulations, on the other hand, we commonly
calculate the one-dimensional free energy profile
$\Delta G(s)\!\propto\! \ln P(s)$ along the pulling coordinate $s$.
To derive an expression for the reaction coordinate probability $P(\vx)$ 
from TMD, we first consider the quasi-static limit adopted in thermodynamic
integration [Eq.\ (\ref{eq:TI})], which conducts
an equilibrium simulation for each value of $s=s(\vq)$.
In direct analogy to Eq.\ (\ref{eq:Px}), we obtain\cite{note1}

\begin{align}\label{eq:cond_prob}
P(\vx,s) &=Q_{\rm eq}^{-1} \upint \!\! \d \vq \d \vp \;
\e^{-\beta H(\vq,\vp) } \,\delta(\vx\!-\!\vx(\vq))\,\delta(s\!-\!s(\vq))
\nonumber \\ 
&=P(s) P(\vx |s) \equiv P(s) \left\langle
  \delta(\vx\!-\!\vx(\vq|s))\right\rangle_{{\rm eq}} ,
\end{align}
where the conditional probability $P(\vx |s)$ represents the
distribution of $\vx$ for a given $s$. Likewise, $\vx(\vq|s)$
represents the collective variable $\vx(\vq)$ restricted to a given
value of $s$. Integration over $s$ readily yields the desired
probability density of coordinate $\vx$,
\begin{equation} \label{eq:prob}
P(\vx)= \upint \!\! \d s \, P(\vx |s) P(s)   
= \upint \!\! \d s \, P(\vx,s).
\end{equation}
By multiplying $P(\vx |s)$ with the TMD weighting $P(s)$, the
distribution $P(\vx)$ and associated free energy
$\Delta G(\vx)\!\propto\! \ln P(\vx)$ represent the correct
equilibrium results.

The situation becomes more involved, if we consider an explicitly
time-dependent Hamiltonian $H (\vq,\vp,t)$. In TMD simulations, for
example, $s(t)\!=\! s_0 \!+ \!v_{\rm c} t$ accounts for moving
distance constraints, with $v_{\rm c}$ denoting the constant pulling
velocity. In other words, the pulling coordinate $s \propto t$
directly corresponds to the time-dependent control parameter in
constrained TMD simulations. As a consequence of the external driving,
the resulting nonequilibrium phase-space density will deviate from a
Boltzmann equilibrium distribution and Eq.~\eqref{eq:cond_prob} does
not hold any more.  Hence we want to resort to a nonequilibrium
formulation, such as Jarzynski's identity in Eq.\
(\ref{eq:Jarzynski}). In fact, Hummer and
Szabo\cite{Hummer01a,Hummer05a} showed that Jarzynski's formulation
can be extended to calculate equilibrium averages of any phase-space
function from a set of nonequilibrium trajectories.

To show this, we employ Jarzynski's identity,
$\left\langle \e^{-\beta W(s)} \right\rangle_{\rm neq} = \e^{-\beta
  \Delta G(s)} \stackrel{{\rm Eq}.\ \!\! (\ref{eq:pmf})}{=}P(s)/P(s_0)$, and
express the nonequilibrium average as an integral over all trajectories
starting from Boltzmann-weighted initial conditions
$(\vq_0, \vp_0, s_0)$,
\begin{align}\label{eq:JI2}
  \frac{P(s)}{P(s_0)} 
&=  \left\langle \e^{-\beta W(s)} \right\rangle_{\rm neq}
  \nonumber \\    
&= Q_{s_0}^{-1} \upint \!\! \d \vq_0 \d \vp_0 \,  \e^{-\beta H(\vq_0, \vp_0, s_0)} 
    \e^{-\beta W(s)},
\end{align}
where
$Q_{s_0}=\upint \!\! \d \vq_0 \d \vp_0 \; \e^{-\beta H(\vq_0, \vp_0,
  s_0)}$.  By inserting $\delta$-functions in the definition of $P(s)$
(analogous to Eq.~\eqref{eq:Px}) and the nonequilibrium average, we
obtain the joint probability
\begin{align}\label{eq:JI3}
\frac{P(\vx,s)}{P(s_0)} &= \left\langle \delta (\vx\!-\!\vx(\vq|s))\,
                          \e^{-\beta W(s)} \right\rangle_{\rm neq}, 
\end{align}
from which the reaction coordinate probability $P(\vx)$ is obtained
via Eq.\ (\ref{eq:prob}). In this way, the equilibrium free energy
landscape 
\begin{align}\label{eq:dGxneq}
\beta \Delta G(\vx) = -  \ln \left[ \frac{\upint \!\d s \left\langle
  \delta (\vx\!-\!\vx(\vq|s))\, \e^{-\beta 
         W(s)} \right\rangle_{\rm neq}}
         {\upint \!\d s'\,\left\langle \e^{-\beta
             W(s')} \right\rangle_{\rm neq}} \right]
\end{align}
can be directly calculated from nonequilibrium TMD simulations.

The above derivation is readily generalized to obtain
equilibrium averages of some phase-space function $A$ via
\cite{Hummer01a,Hummer05a,Crooks00} 
\begin{equation}\label{eq:Asneq}
\left\langle A(\vq|s) \right\rangle_{\rm eq} \frac{P(s)}{P(s_0)}
  =   \left\langle A(\vq|s) \, \e^{-\beta W(s)} \right\rangle_{\rm neq},
\end{equation}
which leads to 
\begin{align}\label{eq:Aneq}
\left\langle A \right\rangle_{\rm eq} &= 
\upint \! \d s\,P(s) \left\langle A (\vq|s)\right\rangle_{\rm eq} \nonumber\\
&= \frac{\upint \!\d s\, \left\langle A(\vq|s)\, \e^{-\beta W(s)}
  \right\rangle_{\rm neq}}{\upint \!\d s'\,\left\langle \e^{-\beta
  W(s')} \right\rangle_{\rm neq}}, 
\end{align}
where the normalization factor $P(s_0)$ is obtained by integrating
Eq.~\eqref{eq:JI2} over $s$.
While this formulation is in principle exact, its practical use
depends on how well observable $A$ is sampled by nonequilibrium
simulations along pulling coordinate $s$. In particular, this includes
the sampling of rare events that affect the estimation of $P(s)$ and
$ \left\langle A (\vq|s) \right\rangle_{\rm eq}$.

Instead of reweighting the nonequilibrium data to obtain equilibrium
averages, it may be advantageous to focus on the nonequilibrium
distribution generated by the TMD simulations pulling in the total
range $\Delta s = s_{\rm max}-s_{\rm min}$,
\begin{align}\label{eq:Pneq}
{\cal P}(\vx) = \frac{1}{\Delta s}  \upint_{s_{\rm
  min}}^{s_{\rm max}} \!\d s\, \left\langle \delta 
  (\vx\!-\!\vx(\vq|s)) \right\rangle_{\rm neq} \, , 
\end{align}
which provides equal weighting of all data points (in contrast to the
equilibrium probability density). This allows us to define the
corresponding ``nonequilibrium energy landscape'' 
\begin{align}\label{eq:dGneq}
\beta \Delta {\cal G}(\vx) = -  \ln \left[\frac{1}{\Delta
  s} \upint_{s_{\rm min}}^{s_{\rm
  max}} \!\!\!\d s\, \left\langle  \delta (\vx\!-\!\vx(\vq|s)) 
  \right\rangle_{\rm neq} \right] \! .
\end{align}
To avoid confusion, we refrain to refer to $\Delta {\cal G}$ as
``nonequilibrium {\em free} energy,'' although this term is used in
information theory.\cite{Parrondo15}

%
%
\subsection{Principal component analysis} \label{sec:TheoPCA}

As explained in the Introduction, PCA is a popular method to construct
low-dimensional reaction coordinates $\vx$, which can be used to
represent the free energy landscape $\Delta G(\vx)$. While the
procedure is straightforward to apply to equilibrium simulations,
several possibilities exist in the nonequilibrium case. To introduce
the basic idea, we first consider the case of an unbiased equilibrium
MD simulation with coordinates $\vq = \{q_i\}$ and the covariance
matrix
\begin{equation}\label{eq:CovEq}
\sigma_{ij} = \mean{ \delta q_i \delta q_j }_{\rm eq},
\end{equation}
where $\delta q_i \!=\! q_i \!-\! \mean{q_i}_{\rm eq}$. PCA
represents a linear transformation that diagonalizes $\sigma$ and thus
removes the instantaneous linear correlations among the
variables. Ordering the eigenvalues of eigenvectors
$\vec{e}_k^{\rm eq}$ decreasingly, the first principal components
\begin{equation}
V_{k}^{\rm eq} = \vec{e}_k^{\rm eq} \cdot  \vec{\delta q} 
\end{equation}
account for the directions of largest variance of the data, and are
therefore often used as reaction
coordinates.\cite{Noe17,Sittel18,Amadei93,Mu05,Altis08} 

We next consider TMD simulations in the quasi-static limit [Eq.\
(\ref{eq:TI})], which conduct an equilibrium simulation for each
value of $s$. In obvious generalization of Eq.\ (\ref{eq:CovEq}), we
define an $s$-dependent covariance matrix
\begin{equation}\label{eq:CovEqS}
\sigma_{ij}(s) = \mean{ \delta q_i(s) \delta q_j(s) }_{\rm eq} ,
\end{equation}
where again
$\delta q_i(s) \!=\! q_i(s) \!-\! \mean{q_i}_{\rm eq}$.\cite{note2}
Averaging over $s$ results in
\begin{equation}\label{eq:CovAv}
\sigma_{ij} = \upint\!\! \d s \, P(s)\, \sigma_{ij}(s).
\end{equation}
Assuming that the correct equilibrium weighting $P(s)$ is used (and
that the constrained simulations are converged), this covariance
matrix is equivalent to the equilibrium result in Eq.\
(\ref{eq:CovEq}), and therefore also yields the same eigenvectors
$\vec{e}_k^{\rm eq}$. In a second step, we calculate the conditional
probability $P(\vx |s)$ from the constrained simulations, using
$x_k \!=\! \vec{e}_k^{\rm eq} \!\cdot\! \vec{\delta q}$. By
averaging $P(\vx |s)$ over 
$s$ with the correct weighting $P(s)$, we
obtain reaction coordinate probability $P(\vx)$ and thus the desired
equilibrium free energy landscape $\Delta G(\vx)$.
We note that the above procedure uses the weighting $P(s)$ of the
constrained simulations twice: First to calculate the equilibrium
covariance matrix from the conditional covariance matrix [Eq.\
(\ref{eq:CovAv})], and second to calculate the equilibrium
distribution $P(\vx)$ from the conditional probability $P(\vx |s)$
[Eq.\ (\ref{eq:prob})]. The former results in adjusted
principal components which represent the data, the latter
corresponds to a reweighting of the data itself.

The above considerations are readily extended to the case of
general time-dependent pulling by replacing Eq.\ (\ref{eq:CovEqS}) by
the Jarzynski-type relation (\ref{eq:Asneq}), yielding
\begin{align}\label{eq:CovEqJS}
\sigma_{ij}(s) &= \frac{ \mean{\delta q_i(s) \delta q_j(s) \,
\e^{-\beta W(s)} }_{\rm neq}}{\mean{ \e^{-\beta W(s)}}_{\rm neq}} . 
\end{align}
Combined with Eq.\ (\ref{eq:CovAv}), we get
\begin{align}\label{eq:CovEqJ}
\sigma_{ij} &= \frac{\upint\!\! \d s \mean{\delta q_i(s) \delta q_j(s) \,
              \e^{-\beta W(s)} }_{\rm neq}}{\upint\!\! \d s^\prime
              \mean{ \e^{-\beta W(s^\prime)} }_{\rm neq}} \, .
\end{align}
Alternatively, it may be desirable to only reweight the data, but use
equally weighted covariances. As discussed above [Eq.\
(\ref{eq:Pneq})], this leads to
\begin{equation}\label{eq:CovAvNEQ}
\sigma^{\rm neq}_{ij} = \frac{1}{\Delta s}
\upint_{s_{\rm min}}^{s_{\rm max}} \!\! \d s \,  
  \mean{ \delta q_i(s) \delta q_j(s) }_{{\rm neq}},
\end{equation}
where fluctuations $\delta q_i$ are referenced with respect to the
mean of the concatenated data. The covariance matrix results in
principal components $V^{\rm neq}_{i}$ that map out an energy
landscape associated with the nonequilibrium distribution generated by
TMD.

As a last --arguably most straightforward-- possibility, we may
refrain from any reweighting and use nonequilibrium principal
components $V^{\rm neq}_{i}$ to directly represent the nonequilibrium
data via $\Delta \mathcal{G}$. In practice, this simply means to
perform a PCA of the concatenated TMD trajectories.  While that
approach may seem somewhat {\em ad hoc} at first sight, it is in fact
well defined, since we know from the above discussion how
nonequilibrium principal components and nonequilibrium data are
connected to their equilibrium counterparts.

%
%
\subsection{Computational Methods}

{\bf MD details.} A 20 $\upmu$s long unbiased MD simulation of
decaalanine (Ala$_{10}$) in vacuo was performed using the GROMACS
2016.3 software package\cite{Abraham15} and the CHARMM36 force
field.\cite{CHARMM36} Employing uncharged protonation states for 
terminal residues, Ala$_{10}$ was set in a dodecahedral box with an
image distance of 11 nm. Following steepest decent minimization, the
initial helical structure was equilibrated under $NVT$ at
$T\!=\!293.5$ K for 10 ns using the Bussi thermostat (v-rescale option
in GROMACS)\cite{Bussi07} with a coupling time constant of 0.2 ps. The
integration time step was set to 1 fs, MD frames were saved every
picosecond. Covalent bonds including hydrogen were constrained by the
lincs algorithm,\cite{Hess08a} electrostatics were described by the
particle mesh Ewald (PME) method,\cite{Darden93} using a direct space
cutoff of 1.2 nm. Van der Waals forces were calculated with a cutoff
of 1.2 nm. Visualization of molecular data was performed with VMD
\cite{Humphrey96} and pymol. \cite{PyMOL15}\\

{\bf Targeted Molecular Dynamics.}  TMD simulations of Ala$_{10}$ in
vacuo were performed employing the PULL code as implemented in GROMACS
2016.3, using the ``constraint'' mode based on the SHAKE
algorithm.\cite{Ryckaert77} The pulling coordinate $s$ was chosen as
the distance between N-terminal nitrogen atom and C-terminal carbonyl
oxygen atom. Other choices, such as a linear combination of contact
distances as provided from contact PCA\cite{Ernst15} yielded overall
quite similar results (data not shown). All simulations started from
the equilibrated system structure after an initial 10 ns $NVT$ run,
using the same thermostat scheme as given above. Translation and
rotation of the center of mass were removed 
(``comm-mode angular'' option), to prevent spinning of Ala$_{10}$ due to
pulling of the asymmetric peptide backbone. Two sets of simulations
with constant velocity $v_{\rm c}$ = 1 m/s were performed: $10\, 000$
trajectories from $s=1.1-2.1$ nm and 100 trajectories from $s=1.5-3.5$
nm. Constraint forces were saved each
time step (1 fs), Cartesian coordinates every 0.1 ps.\\

{\bf Dihedral angle principal component analysis.}
Since Cartesian coordinates unavoidably result in a mixing of overall
rotation and internal motion,\cite{Sittel14} internal coordinates are
used for PCA. Here we employ ($\phi_i,\psi_i$) backbone dihedral 
angles, which have been shown to be well suited to describe the
dynamics of peptides and small proteins.\cite{Mu05,Altis07,
  Ernst15,Sittel18} To take the periodicity of the dihedral angles
into account, we shift the periodic boundary of the circular data to
the region of the lowest point density. This ``maximal gap shifting''
approach was incorporated into the new version of the dihedral angle
principal component analysis (dPCA+),\cite{Sittel17} which represents
a significant improvement to the previously advocated
sine/cosine-transformed variables used in dPCA.\cite{Mu05,Altis07} It
avoids artificial doubling of coordinates and distortion errors due to
the nonlinearity of the sine and cosine transformations.
In the case of Ala$_{10}$, dPCA+ was performed on the
($\phi_i,\psi_i$) dihedral angles of the eight inner residues. While
the first six equilibrium principal components show multipeaked
distributions, the first two components already cover $\approx$ 70 \%
of the overall variance (Fig.\ \SIPCAEQ a,b). Since the maximal gap shifts
may differ for unbiased and biased data (Fig.\ \SIPCAEQ c), for
consistency we used in all cases the shifts obtained from the
reweighted nonequilibrium data (which are equivalent to shifts
obtained from the equilibrium
simulation).\\

{\bf Free energy calculations.}
To evaluate the free energy landscape $\Delta G(\vx)$ via the Jarzynski-type
expression in Eq.\ (\ref{eq:dGxneq}), the probability density
$P(\vec{\CV})$ is estimated from the TMD data by a weighted
histogram. Defining $\delta_{k,\CC}(\vec{\CV})$ as a counting function
in some bin size $\Delta\vec{\CV}$,
\begin{equation}
\delta_{k,\CC}(\vec{\CV}) = \left\{ \begin{array}{l}
1 \quad \text{if} \;\; \vec{\CV}-\tfrac{\Delta\vx}{2} \geq \;
 \vec{\CV}_k(\CC) \; >\vec{\CV}+\tfrac{\Delta\vx}{2} \\ 
0 \quad \text{else} \; ,
\end{array} \right.
\end{equation}
and utilizing all concatenated trajectories with in total $N$ data
points, the estimator of the free energy reads
\begin{equation} \label{eq:dGcalc}
\beta \Delta G(\vec{\CV}) = - \ln \left[ \frac{\sum_{n}^{N}
    \euler^{-\beta W_n} \, \delta_n (\vec{\CV})}{\sum_{n}^{N}
    \euler^{-\beta W_n}}  \right]  \; . 
\end{equation}
In a similar way, the expectation value of a general observable $A$ is
estimated via
\begin{equation}
\mean{A} = \frac{\sum_{n}^{N} \euler^{-\beta W_n} \,
  A_n}{\sum_{n}^{N} \euler^{-\beta W_n}}   \; .
\end{equation}
Using Gaussian smoothing, we avoid sharp edges in the histogram due to
low-work trajectories contributing to almost empty bins.

Apart from direct evaluation of Jarzynski's identity via Eq.\
(\ref{eq:dGcalc}), we also consider the recently proposed dissipation
corrected TMD approach.\cite{Wolf18} Employing Langevin theory, the
dissipated energy $W_{\rm diss}= \mean{W}-\Delta G$ can be expressed as
\begin{align}\label{eq:dcdG}
W_{\rm diss} (s) = v_{\rm c} \upint_{s_0}^{s}\mathrm{d}s'
        \,  \itGamma(s') , 
\end{align}
where $\itGamma(s)$ represents the position-dependent friction
coefficient of the system. Using second-order cumulant approximation,
this friction is estimated as\cite{Wolf18}
\begin{align}\label{eq:Gamma}
\itGamma(s) =  \frac{\beta}{v_{\rm c}} \upint_{s_0}^{s} \mathrm{d}s' \,
\left\langle \delta f_{\rm c}(s) \delta f_{\rm c}(s')
  \right\rangle_{\rm neq} ,   
\end{align}
which may be calculated on-the-fly from
constraint force fluctuations $\delta f_{\rm c}(s) = f_{\rm c}(s)
- \left \langle {f}_\text{c}(s) \right \rangle_{\rm neq}$. In cases where
the underlying assumption of a Gaussian work distribution is roughly
fulfilled, Eq.\ (\ref{eq:dcdG}) was shown to converge significantly
faster than the direct evaluation [Eq.\ (\ref{eq:Jarzynski})]. As a
bonus, the approach also provides the friction profile $\itGamma(s)$,
which presents a microscopic picture of the system-bath coupling.

%
%
\section{Results and Discussion}

To investigate the applicability of the above developed formulation,
we adopt Ala$_{10}$ in vacuo, which has been used by several groups to
study the enforced unfolding of the $\alpha$-helix.\cite{Park03,
  Procacci06,Forney08,Oberhofer09,Hazel14} This process, though,
virtually does not occur in equilibrium simulations at room
temperature. Vice versa, unbiased MD is found to sample conformational
states that are difficult to come by with TMD. By comparing unbiased
MD and nonequilibrium TMD simulations in several regimes of the
pulling coordinate $s$, in the following we study the virtues and
shortcomings of TMD and discuss PCA of nonequilibrium data.

%
%
\subsection{Unbiased MD simulations}

We first consider the 20~$\upmu$s long unbiased MD trajectory of
Ala$_{10}$ in vacuo, which was analyzed using dPCA+ (see Methods).
Figure \ref{fig:fe_eq}a shows the resulting free energy landscape
along the first two principal components $V_1^{\rm eq}$ and
$V_2^{\rm eq}$, which represent about 70 \% of the overall variance
of the system. The energy landscape clearly reveals the main
metastable conformational states of Ala$_{10}$, including a
hairpin-like conformation {\em HP} (populated by 39 \%), the
$\alpha$-helix $A$ (23 \%) and a helical state $B$ (20 \%) that is
broken at the C-terminus. Moreover we find several ``pretzel-shaped''
conformations, here termed $P$ (2 \%), $C$ (3 \%) and {\em CH} (2 \%),
while extended conformations $U$ are not sampled in unbiased MD.

\begin{figure}
\begin{centering}	
\includegraphics[width=8cm]{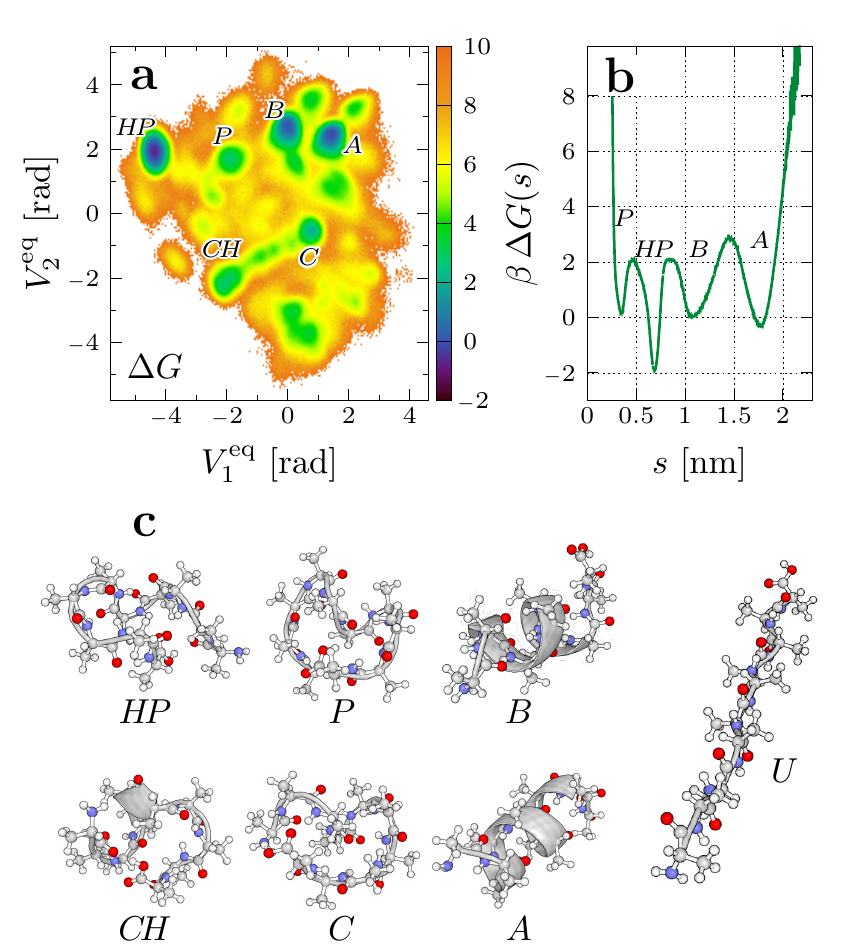}
\caption{
Unbiased MD simulation of Ala$_{10}$. (a) Free energy landscape
$\Delta G(V_1^{\rm eq},V_2^{\rm eq})$ (in units of $\kb T$) along the
first two principal components obtained from dPCA+. (b) Free energy
profile $\Delta G(s)$ with respect to the pulling coordinate. (c)
Molecular structures of the main metastable conformational states.} 
\label{fig:fe_eq}
\end{centering}
\end{figure}

Projecting the unbiased data onto pulling coordinate $s$ (which here
represents an unconstrained stochastic variable), the free energy
profile $\Delta G(s)$ only reveals the main conformational states {\em
  HP}, $A$, $B$ and $P$ (Fig.\ \ref{fig:fe_eq}b). In particular, we
note that the connectivity is not preserved in the one-dimensional
representation, since state {\em HP} (instead of $P$) is now direct
neighbor of state $B$. In fact, when we plot the free energy as a
function of $s$ and $V_1^{\rm eq}$ or $V_2^{\rm eq}$ (Fig.\ \SIFELsV), we
find that for $s\lesssim 1$~nm several conformational states may
coexist for the same value of $s$. As a consequence, the time
evolution of $s(t)$ exhibits jumps between $s \approx$ 0.4 and 1.1 nm
(Fig.\ \SIFELsV), reflecting that the system directly transits from $P$ to
$B$ (as suggested by Fig.\ \ref{fig:fe_eq}a). Hence for
$s\lesssim 1$~nm the pulling coordinate $s$ represents a poor choice
of a reaction coordinate.

%
%
\subsection{Comparison of unbiased and constrained simulations}

The discussion above indicates that TMD simulations are difficult to
interpret for $s\lesssim 1$~nm, since several free energy minima may
occur for the same value of pulling coordinate $s$. On the other hand,
we noticed that the sampling of the unbiased simulation is restricted
to $s\!\lesssim \!2$~nm (Fig.\ \ref{fig:fe_eq}b), although TMD
simulations may be extended to study the unfolding of Ala$_{10}$
($s\!\approx \!3$~nm, see below). To achieve a meaningful comparison
of MD and TMD simulations, in the following we therefore restrict
ourselves to the range of 1.1~nm$\,\le s\le\,$2.1~nm, which enables us
to describe transitions between states $B$ and $A$. In particular, the
comparison allows us to validate the theory developed above. 

\begin{figure}[h]
\begin{centering}
\includegraphics{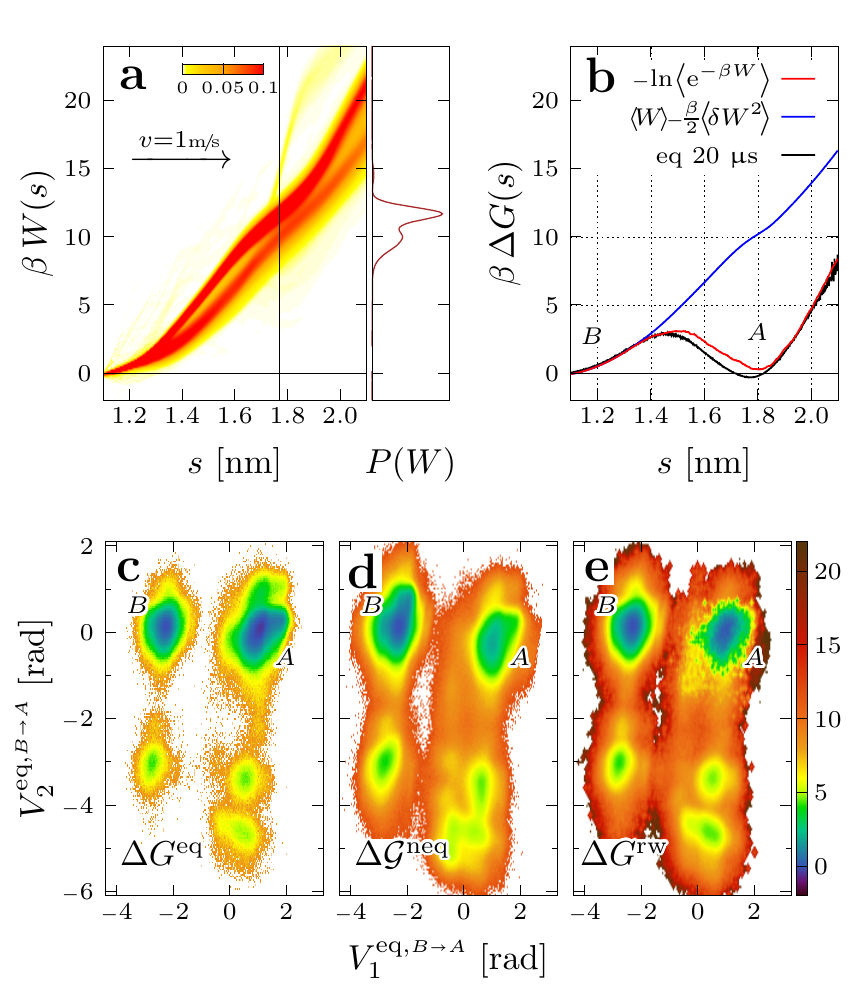}
\caption{
  Comparison of unbiased MD and nonequilibrium TMD simulations for the
  interval 1.1~nm$\,\le s\le\,$2.1~nm, describing transitions between
  states $B$ and $A$ of Ala$_{10}$.  (a) Distribution of work
  performed on the system by external pulling along coordinate $s$. On
  the right, the cut $P(W)$ at $\CC=1.77$ nm reveals non-Gaussian
  structure. (b) Comparison of free energy profiles $\Delta G(s)$
  obtained from unbiased MD simulation (black), Jarzynski's identity
  (red), and second-order cumulant approximation (blue). (c-e) Energy
  landscapes (in units of $\kb T$) as a function of the first two
  principal components obtained from dPCA+ performed for
  1.1~nm$\,\le s\le\,$2.1~nm.  Compared are (c) results from unbiased
  MD, (d) TMD, and (e) reweighted TMD.}
\label{fig:fe_val}
\end{centering}
\end{figure}
  
In order to characterize the nonequilibrium simulations, we recall
that $\Delta G = \mean{W}- W_{\rm diss}$, stating that the free energy
difference results from the work performed on the system minus the
dissipated energy. To begin with the performed work, Fig.\
\ref{fig:fe_val}a shows that the work distribution reveals a
complicated structure, including two prominent maxima and several
smaller contributions due to rare and wide-spread trajectories.  The
associated free energy profile $\Delta G(s)$ obtained from Jarzynski's
identity (Fig.\ \ref{fig:fe_val}b) shows two minima reflecting states
$B$ and $A$. This result agrees well with the outcome of the unbiased
simulations, while the second-order cumulant approximation [Eq.\
(\ref{eq:dcdG})] fails to reproduce $\Delta G(s)$ due to the
non-Gaussian structure of the work distribution. Owing to the
complicated structure of the work distribution, we needed to run
$10\, 000$ short nonequilibrium trajectories to achieve satisfactory
agreement of TMD and unbiased simulations (see Fig.\ \SIConv a,b for a
study of the convergence behavior). This is a consequence of the
exponential average,
$\e^{-\beta \Delta G(s)} = \left\langle e^{-\beta W(s)}
\right\rangle_{\rm neq}$, where mainly rare low-$W$ trajectories
dominate the free energy estimate. Since $\Delta G(s)$ is considerably
lower than the average work $\langle W(s) \rangle$, the stretching of
state $B$ into helix $A$ with velocity $v_{\rm c}=1$ m/s generates
considerable irreversible heat via intramolecular
friction.\cite{Cellmer08,Schulz12,Soranno12,Erbas13,Echeverria14}

Having verified that the TMD simulations correctly reproduce the free
energy profile $\Delta G(s)$, we are in a position to consider to what
extent TMD allows us to predict the free energy along a general
reaction coordinate $\vx$. As suitable coordinates we choose the first
two principal components
$V_1^{\text{eq},\scriptscriptstyle{B\shortrightarrow A}}$ and
$V_2^{\text{eq},\scriptscriptstyle{B\shortrightarrow A}}$ obtained
from dPCA+, which was performed for all unbiased trajectory points
that lie in the interval 1.1~nm$\,\le s\le\,$2.1~nm.  Figure
\ref{fig:fe_val}c shows the resulting free energy landscape obtained
from unbiased MD data. Compared to the energy landscape pertaining to
the complete data set (Fig.\ \ref{fig:fe_eq}a), we note that only the
adjacent minima of $B$ and $A$ are included, since all other states
are associated with values of $s \le 1.1\,$nm.
The two-dimensional representation
$\Delta G(V_1^{\text{eq},\scriptscriptstyle{B\shortrightarrow
    A}},V_2^{\text{eq},\scriptscriptstyle{B\shortrightarrow A}})$ can
be employed to explain the prominent features of the work distribution
(Fig.\ \ref{fig:fe_val}a) in terms of pathways on the free energy
surface. Roughly speaking, high-$W$ trajectories mostly transfer
directly between states $B$ and $A$, while low-$W$ trajectories
typically do not reach state $A$ at $s=2.1\,$nm, since several
populated regions coexist for this value of $s$ (Fig.\ \SIFELsVneq).
We note that in general there is no direct correspondence between
routes in work space and paths in real space.

Using the same coordinates, Fig.\ \ref{fig:fe_val}d shows the energy
landscape associated with the nonequilibrium distribution generated by
the TMD simulations [Eq.\ (\ref{eq:dGneq})]. Overall, nonequilibrium
results and unbiased results (Fig.\ \ref{fig:fe_val}c) appear quite
similar, because the free energy $\Delta G(s)$ (and thus the weighting
$P(s)$) pertaining to states $B$ and $A$ is alike. In detail, however,
the nonequilibrium energy landscape shows a population shift from
state $A$ to some side minima at lower values of
$V_2^{\text{eq},\scriptscriptstyle{B\shortrightarrow A}}$. Moreover,
the TMD simulations affect a sampling of high-energy regions (shown in
orange), that are not accessible to the unbiased simulation.
Lastly, Fig.\ \ref{fig:fe_val}e shows the energy landscape associated
with the reweighted nonequilibrium data [Eq.\ (\ref{eq:dGxneq})]. As
expected, this energy landscape is indeed quite similar to the
unbiased equilibrium result in Fig.\ \ref{fig:fe_val}c. Considering
the high amount of dissipated work, this similarity appears quite
remarkable.

To compare equilibrium, nonequilibrium, and reweighted nonequilibrium
data (Figs.\ \ref{fig:fe_val}c-e), we have so far employed principal
components generated from unbiased equilibrium MD. Alternatively,
these data may be also examined using principal components generated
from nonequilibrium data, see Eq.\ (\ref{eq:CovAvNEQ}).  Owing
to the similar weighting $P(s)$ of states $B$ and $A$, the resulting
energy landscapes (Fig.\ \SIFELneqPC a) are again quite similar and
hardly yield new information.
The difference between principal components generated from equilibrium
or nonequilibrium data can also be directly studied by comparing the
respective covariance matrices. Since the $B\!\rightarrow\!A$
transition mainly involves the folding of the C-terminus residues,
TMD simulations that enforce this transition are found to result in
enhanced correlations between the last three residues (Fig.\
\SIFELneqPC b). Upon reweighting, the covariance matrix again resumes the
structure of the unbiased equilibrium MD.

To summarize, we have shown that the $B\!\rightarrow\!A$ transition of
Ala$_{10}$ can be viewed using principal components generated from
equilibrium data [Eq.\ (\ref{eq:CovAv})] or nonequilibrium data [Eq.\
(\ref{eq:CovAvNEQ})]. Both representations are well defined as they
are simply related via the weighting function $P(s)$. Independent of
this choice of representation, we may consider equilibrium,
nonequilibrium, or reweighted nonequilibrium data to represent the
energy landscapes of the system, see Figs.\ \ref{fig:fe_val}c-e. Due
to the similar weighting $P(s)$ of states $B$ and $A$, so far the
resulting energy landscapes exhibited only minor differences (but see
below).

%
%
\subsection{TMD simulation of helix unfolding}

As a well-established application of pulling simulations,
\cite{Park03,Procacci06,Forney08,Oberhofer09,Hazel14} we consider in
Fig.\ \ref{fig:fe_unf} the unfolding of the $\alpha$-helical state of
Ala$_{10}$. Since the free energy difference between helical state $A$
and extended state $U$ is quite large ($\approx 28\, k_{\rm B}T$),
this process does virtually not occur in the 20 $\upmu$s long unbiased
MD trajectory which only samples up to $s \lesssim 2.2\,$nm. In our
TMD simulations, all trajectories start at $s=1.5$ nm in
$\alpha$-helical structure $A$, run into a local energy minimum
(corresponding to a more favorable helical structure), and
successively unfold until they reach the extended state $U$ at
$s\approx 3.1$ nm. Unlike the case of the above studied
$B\!\rightarrow\!A$ transition, the work distribution of the
$A\!\rightarrow\!U$ transition is mono-modal and well approximated by
a Gaussian (Fig.\ \ref{fig:fe_unf}a). As a consequence, the free
energy profile $\Delta G(s)$ obtained from Jarzynski's identity and of
its second-order cumulant approximation [Eq.\ (\ref{eq:dcdG})] are in
perfect agreement (Fig.\ \ref{fig:fe_unf}b). Moreover, we find that
the free energy rapidly converges for already 100 TMD runs (Fig.\
\SIConv c). This is a consequence of the fact that states $A$ and $U$
are connected by only two well-defined and well-accessible paths that
require a minimum number of contact changes (see below).

\begin{figure}
\begin{centering}	
\includegraphics[width=8cm]{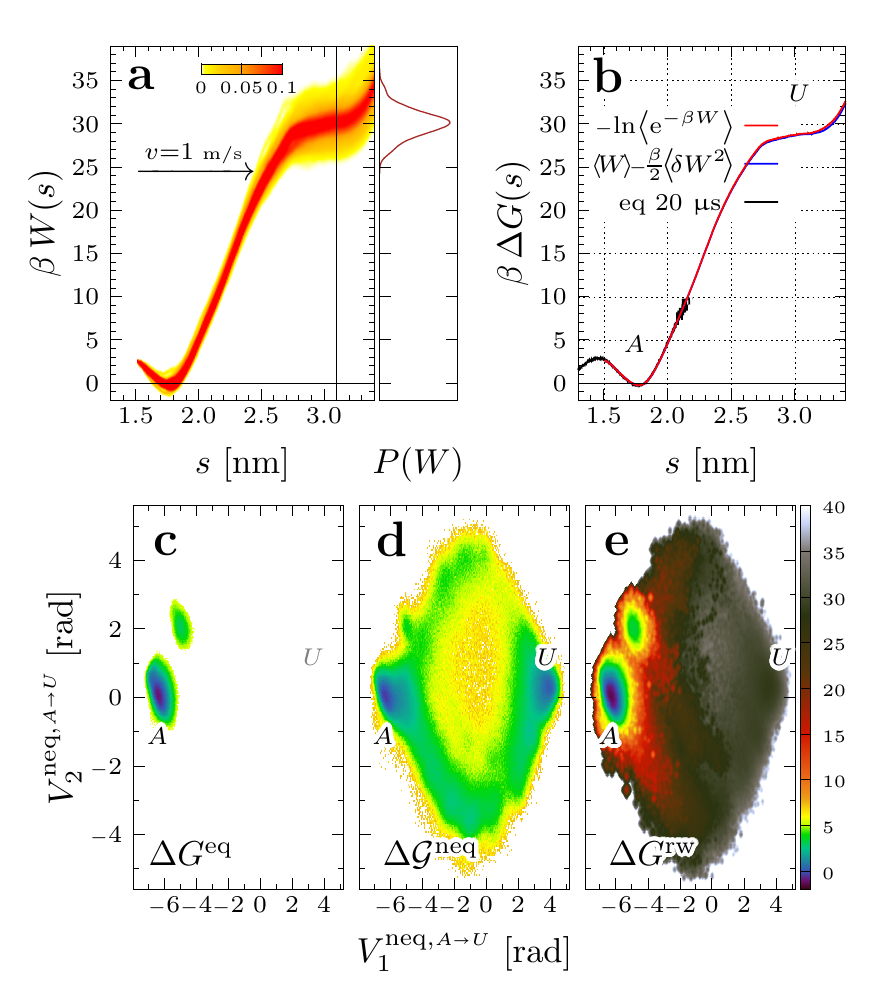}
\caption{
  Unfolding of the $\alpha$-helical state $A$ of Ala$_{10}$. (a)
  Distribution of work performed on the system by external pulling
  along coordinate $s$, and a cut $P(W)$ at $\CC=3.1$ nm. (b)
  Comparison of free energy profiles $\Delta G(s)$ obtained from
  unbiased MD simulation (black), Jarzynski's identity (red), and
  second-order cumulant approximation (blue). (c-e) Energy landscapes
  (in units of $\kb T$) as a function of the first two principal
  components obtained from dPCA+ performed for
  1.5~nm$\,\le s\le\,$3.5~nm. Compared are (c) results from unbiased
  equilibrium MD, (d) nonequilibrium TMD, and (e) reweighted TMD.}
\label{fig:fe_unf}
\end{centering}
\end{figure}

Due to the large free energy difference of states $A$ and $U$ (and the
associated different weighting $P(s)$), we expect large differences
when we perform a PCA of unbiased equilibrium and constrained
nonequilibrium simulations, respectively. This can be illustrated by
the associated covariance matrices which are compared in Fig.\
\ref{fig:fe_Cov}a. While in the equilibrium case (using all data with
$s>1.5$ nm) [Eq.\ (\ref{eq:CovEq})] we find moderate correlations of
mostly neighboring residues, virtually all residues are correlated in
the nonequilibrium covariance matrix [Eq.\ (\ref{eq:CovAvNEQ})]. This
is a consequence of the fact that upon unfolding all backbone dihedral
angles change from $\alpha$-helical to extended structures. Upon
reweighting the nonequilibrium covariances [Eq.\ (\ref{eq:CovAv})], we
recover the equilibrium result, as expected.

Let us consider the resulting equilibrium and nonequilibrium principal
components $V_k^{\text{eq},\scriptscriptstyle{A\shortrightarrow U}}$
and $V_k^{\text{neq},\scriptscriptstyle{A\shortrightarrow U}}$,
respectively.  To elucidate which coordinates are better suited to
describe the $A\!\rightarrow\!U$ unfolding process, it is instructive
to study the eigenvectors pertaining to the first two components,
which account for 50 \% (eq) and 90 \% (neq) of the total variance,
respectively. As shown in Fig.\ \ref{fig:fe_Cov}b, the eigenvectors of
equilibrium components
$V_1^{\text{eq},\scriptscriptstyle{A\shortrightarrow U}}$ and
$V_2^{\text{eq},\scriptscriptstyle{A\shortrightarrow U}}$ report
exclusively on local motions at the C- and N-terminus, respectively
(which is mainly what happens at equilibrium). On the other hand, the
eigenvectors of nonequilibrium components
$V_1^{\text{neq},\scriptscriptstyle{A\shortrightarrow U}}$ and
$V_2^{\text{neq},\scriptscriptstyle{A\shortrightarrow U}}$ are found
to account for the global motion of all residues and thus report
directly on the $A\!\rightarrow\!U$ unfolding process.\cite{note3}
As a further illustration, we plot the energy landscape pertaining to
the nonequilibrium data as a function of $s$ and
$V_1^{\text{eq},\scriptscriptstyle{A\shortrightarrow U}}$ or
$V_1^{\text{neq},\scriptscriptstyle{A\shortrightarrow U}}$ (Fig.\
\ref{fig:fe_Cov}c). Since the pulling coordinate evidently corresponds
to the direction of maximal variance, we find a direct correlation
between $s$ and
$V_1^{\text{neq},\scriptscriptstyle{A\shortrightarrow U}}$. The second
component, on the other hand, is found to split up in two pathways
along $s$, thus providing important information beyond the
one-dimensional free energy profile $\Delta G(s)$.

\begin{figure}
\begin{centering}	
\includegraphics[width=8cm]{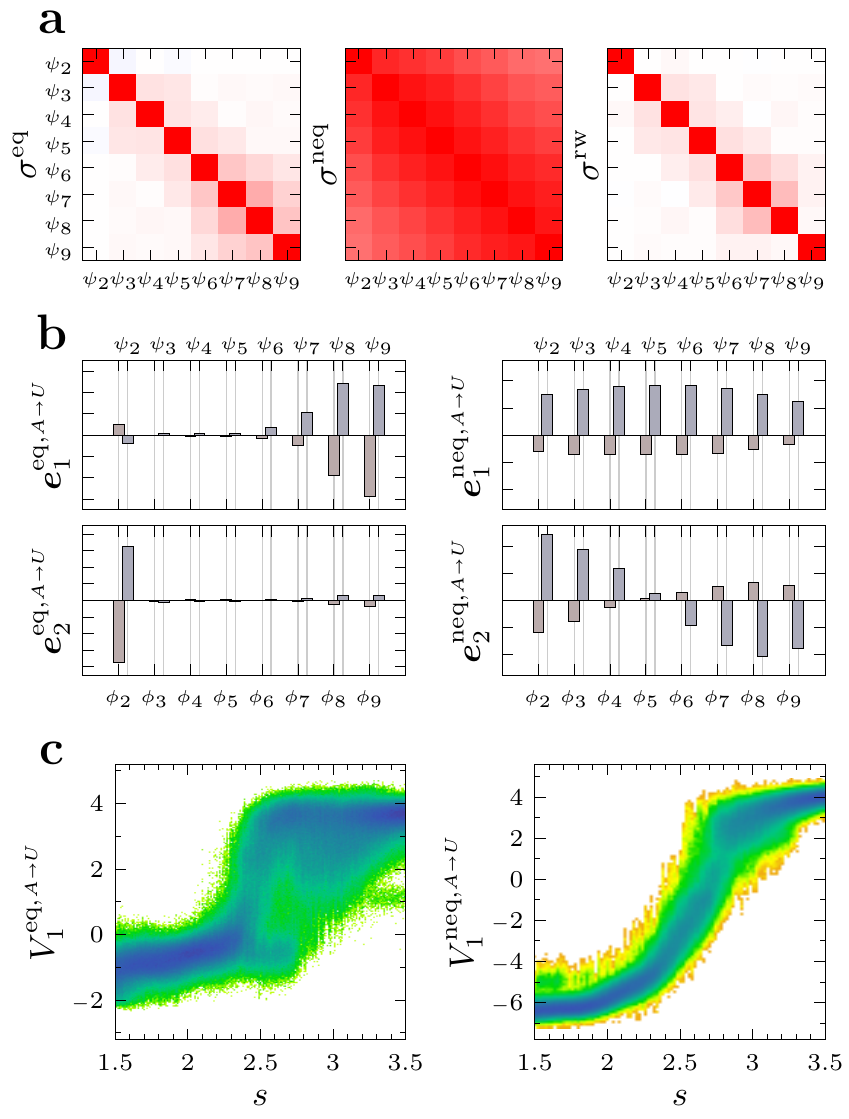}
\caption{
  PCA of the $A\!\rightarrow\!U$ unfolding of Ala$_{10}$. (a)
  Correlation matrices (i.e., normalized covariances) obtained from
  (left) unbiased equilibrium MD, (middle) nonequilibrium TMD, and
  (right) reweighted TMD. (b) Eigenvectors pertaining to the first two
  principal components, obtained from (left) equilibrium MD and
  (right) nonequilibrium TMD. (c) Energy landscapes (in units of
  $\kb T$) as a function of pulling coordinate $s$ and
  $V_1^{\text{eq}, A\shortrightarrow U}$ or
  $V_1^{\text{neq}, A\shortrightarrow U}$.}
\label{fig:fe_Cov}
\end{centering}
\end{figure}

We are now in a position to illustrate the $A\!\rightarrow\!U$
unfolding of Ala$_{10}$ by a multidimensional energy landscape. Using
the first two nonequilibrium principal components
$V_k^{\text{neq},\scriptscriptstyle{A\shortrightarrow U}}$, Fig.\
\ref{fig:fe_unf} shows energy landscapes constructed from (c) unbiased
equilibrium MD, (d) nonequilibrium simulations, and (e) reweighted
nonequilibrium data. As expected, the unbiased free energy landscape
(Fig.\ \ref{fig:fe_unf}c) only samples initial state $A$ together with
a neighboring state that reflects the breaking of the helix at the
N-terminus. The results for the reweighted nonequilibrium data (Fig.\
\ref{fig:fe_unf}e) are quite similar, but also show enhanced sampling
of high-energy regions. Notably, the energy landscape obtained from
the nonequilibrium data (Fig.\ \ref{fig:fe_unf}d) is most informative,
as it shows the entire conformational space sampled by the TMD
simulation including initial state $A$ and final state $U$.

The nonequilibrium energy landscape $\Delta{\cal G}$ [Eq.\
(\ref{eq:dGneq})] indicates two main unfolding pathways, which are
discriminated by the second principal component. The upper half circle
connecting states $A$ and $U$ reflects trajectories that start
unfolding at the N-terminus and continue to the C-terminus, while the
lower half circle corresponds to unfolding trajectories proceeding the
opposite way. Followed by 75 \% of all trajectories, the
C$\rightarrow$N path clearly represents the main unfolding route. This
may be a consequence of the fact that the C-terminus is able to form
hydrogen bonds with both of its oxygen atoms (while the N-terminus can
only form a single hydrogen bond), and therefore exhibits larger
fluctuations and structural destabilization. Trajectories that
initially proceed the opposite N$\rightarrow$C path mostly do not
complete this route, but return to the helical state $A$.

To illustrate the C$\rightarrow$N unfolding pathway, Fig.~\SIDiheNEQ a
shows the evolution of the
peptide's backbone dihedral angles $\psi_n$. As expected, the dihedral
angles change sequentially from an $\alpha$-helical
($\psi \approx -40^\circ$) to an extended ($\psi \approx 160^\circ$)
conformation. Using DSSP\cite{Kabsch83}  to characterize the secondary
structure of Ala$_{10}$, however, we find that the helix does not
unfold directly, but first changes to a 3$_{10}$-helix for
$s \gtrsim 2$ nm (Fig.~\SIDiheNEQ b). In the course of the unfolding
process, the 3$_{10}$-helix may temporarily turn into a shortened
$\alpha$-helix in combination with turn/coil structures. This can
occur anywhere in the peptide sequence, thus allowing the helix to
break at its weakest end.

As a further characterization of the unfolding mechanism, it is
instructive to consider the friction profile $\itGamma(s)$ obtained from
dissipation-corrected TMD\cite{Wolf18} (Fig.~\SIDiheNEQ c).
Reflecting the fluctuations of the constraint force [Eq.\
(\ref{eq:Gamma})], $\itGamma(s)$ is not necessarily related to the form
of the free energy profile $\Delta G(s)$. As in the previously studied
NaCl/water system,\cite{Wolf18} the friction profile may therefore provide new
microscopic information on the unfolding process.
At the onset of unfolding at $s \gtrsim 2.0$ nm, $\itGamma(s)$ starts to
increase and comes to a maximum at full extension at $s \approx 2.8$
nm. We attribute this rise in friction to the loose C-terminal chain,
which can fluctuate more with increasing length. The sharp minimum of
$\itGamma(s)$ at $s \approx 3.0$ nm coincides with a shallow minimum of
the $\Delta G(s)$ profile, pointing to a structural relaxation of the
chain in the extended conformation. For $s \gtrsim 3.0$ nm the
friction increases again, which most likely results
from over-stretching the peptide chain.

%
%
\section{Conclusions}

Aiming to describe nonequilibrium phenomena in terms of a
multidimensional energy landscape, we have studied the application of
dimensionality reduction techniques to nonequilibrium MD data. To be
specific, we have focused on principal component analysis (PCA) of
targeted MD (TMD) simulations\cite{Schlitter93A,
  Schlitter94,Schlitter01} that are constrained along some biasing
coordinate $s$. We have found that it is generally valid to
simply perform PCA on the concatenated nonequilibrium
trajectories. While the resulting distribution ${\cal P}(\vx)$ and energy
landscape $\Delta{\cal G} (\vx) \propto \ln {\cal P}(\vx)$ will not reflect
the equilibrium state of the system, the nonequilibrium energy
landscape may directly reveal the molecular reaction
mechanism. Applied to the unfolding of the $\alpha$-helical state of
Ala$_{10}$, for example, we have identified two unfolding pathways
starting from the C- and N-terminus, respectively. Notably, this
information is not available from the commonly calculated free
energy profile $\Delta G(s)\propto \ln P(s)$.

The nonequilibrium energy landscape $\Delta{\cal G} (\vx)$ is well
defined, because it is related to the equilibrium free energy
landscape through weighting function $P(s)$ accounting for the bias
introduced by TMD. That is, by reweighting the TMD conditional
probability $P(\vx |s)$ by $P(s)$ and subsequently integrating over
$s$ [Eq.\ (\ref{eq:prob})], we obtain the correct equilibrium
distribution $P(\vx)$. The same holds for PCA, where we construct
principal components from nonequilibrium data which are associated to
equilibrium principal components constructed from the reweighted data
[Eq.\ (\ref{eq:CovAv})]. Although this formulation is in principle exact,
its practical use depends on how well the conformational distribution
of interest is sampled by nonequilibrium simulations along biasing
coordinate $s$. Moreover, it is important that coordinate $s$ accounts
for a naturally occurring motion between two well-defined end-states of
the system. This is the case for the example of the
$A\!\rightarrow\!U$ unfolding reaction of Ala$_{10}$ (Fig.\
\ref{fig:fe_unf}), but less so for the $B\!\rightarrow\!A$ transition
(Fig.\ \ref{fig:fe_val}), where the start and end state split
up in various metastable states. 

While we have focused the discussion on TMD, the above described
approach is readily applied to various types of nonequilibrium
simulations. In particular, this implies enhanced sampling methods
that are described by a continuous and sufficiently slow development
along some control parameter $s$ and provide a weighting function
$P(s)$, such as umbrella sampling\cite{Torrie77} and steered
MD,\cite{Isralewitz01,Park04} conformational
flooding,\cite{Grubmueller95} metadynamics,\cite{Laio02} and adaptive
biasing force sampling.\cite{Darve08} Moreover, besides PCA
alternative dimensionality reduction techniques may be employed
including nonlinear techniques\cite{Rohrdanz13,Duan13} and various
kinds of machine learning approaches. \cite{Galvelis17,
  Chen18,Ribeiro18,Brandt18} In ongoing work, we use nonequilibrium
PCA to study conformational changes of T4 lysozyme,\cite{Ernst17} and
to analyze unbinding simulations of small organic molecules from
proteins such as the N-terminal domain of Hsp90\cite{Amaral17} and the
$\beta_2$ adrenergic receptor.\cite{Cherezov07}

%
%
\subsection*{Supplementary Material}
\vspace*{-4mm} Details of dPCA+, energy landscapes as a function of
$s$ and various principal components, evolution of pulling coordinate,
dihedral angles, secondary structure content and 
friction content, convergence tests of free energy estimators.

\subsection*{Acknowledgment}
\vspace*{-4mm} We thank Simon Bray for instructive and helpful
discussions. This work has been supported by the Deutsche
Forschungsgemeinschaft (Sto 247/11) and the bwUniCluster computing
initiatives of the State of Baden-W\"urttemberg. We furthermore
acknowledge support by the High Performance and Cloud Computing Group
at the Zentrum f\"ur Datenverarbeitung of the University of
T\"ubingen, the state of Baden-W\"urttemberg through bwHPC, and the
Deutsche Forschungsgemeinschaft through grant no.\ INST 37/935-1 FUGG.
 
The dPCA+ method\cite{Sittel17} was implemented in the open source
software \emph{FastPCA}. Dissipation-corrected TMD\cite{Wolf18} was
implemented using Python3. All programs are freely available at
\url{https://github.com/moldyn}. 

%
%

\begin{thebibliography}{10}
\bibitem{note1}
{We note in passing that in general the change from free to constrained motion
  involves the inclusion of the so-called Fixman potential, \cite{Fixman74}
  associated with the integration over the conjugated momentum of constrained
  coordinate $s$. Choosing $s$ to be a linear combination of interatomic
  distances (as is usually done in TMD), however, this determinant reduces to a
  constant~\cite{Schlitter01}}.
\bibitem{note2}
{Note that $\delta q_i(s) \!=\! q_i(s) \!-\! \mean{q_i}_{\rm eq}$ is the
  deviation of $q_i$ restricted by a fixed $s$ from its equilibrium mean (not
  the average at $s$), which for TMD simulations can be readily calculated via
  Eq.~\eqref{eq:Aneq}}.
\end{thebibliography}

\end{document}


\title{Principal Component Analysis of nonequilibrium Molecular Dynamics simulations\\
Supplementary Material}
\author{Matthias Post}
\author{Steffen Wolf}
\author{Gerhard Stock}
\affiliation{Biomolecular Dynamics, Institute of Physics, Albert Ludwigs
 University, 79104 Freiburg, Germany}
\thanks{stock@physik.uni-freiburg.de}
\date{21 January 2019}
\maketitle
\centering
%
%

\baselineskip6mm

\newpage

\begin{figure}[h!]
  \centering
    \includegraphics{\dirfig/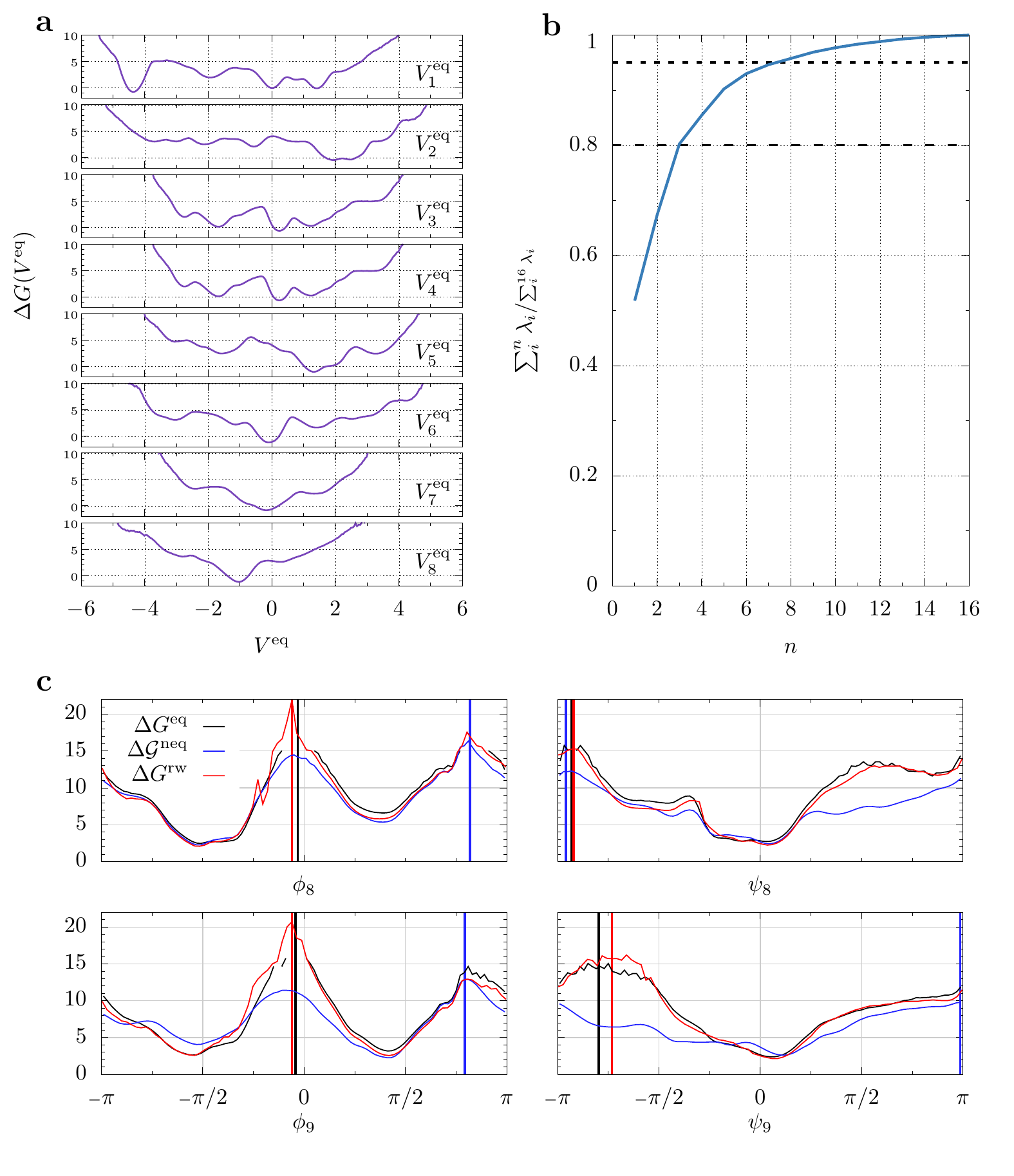}
  \caption{Dihedral angle principal component analysis (\dPCA) of the unbiased MD data of \ala~in vacuo. (a) Free energy projected onto the first equilibrium principle components. (b) Normalized cumulative eigenvalues of the covariance matrix. The first three PCs take 80\% of the variance into account. (c) Backbone dihedral angle distributions of residue 8 and 9 at the C-terminus, obtained from the equilibrium MD data (black), TMD data (blue) and reweighted TMD data (red) of the $B\rightarrow A$ transition. Vertical lines represent the maximal gap determined in the three cases. Due to the enhanced sampling of the TMD data around $\phi\approx 0$, the maximal gap is shifted to $\approx 3\pi/4$ in this case.}
  \label{fig:eq_pcs_fel}
\end{figure}

\begin{figure}[h!]
	\centering
	\includegraphics{\dirfig/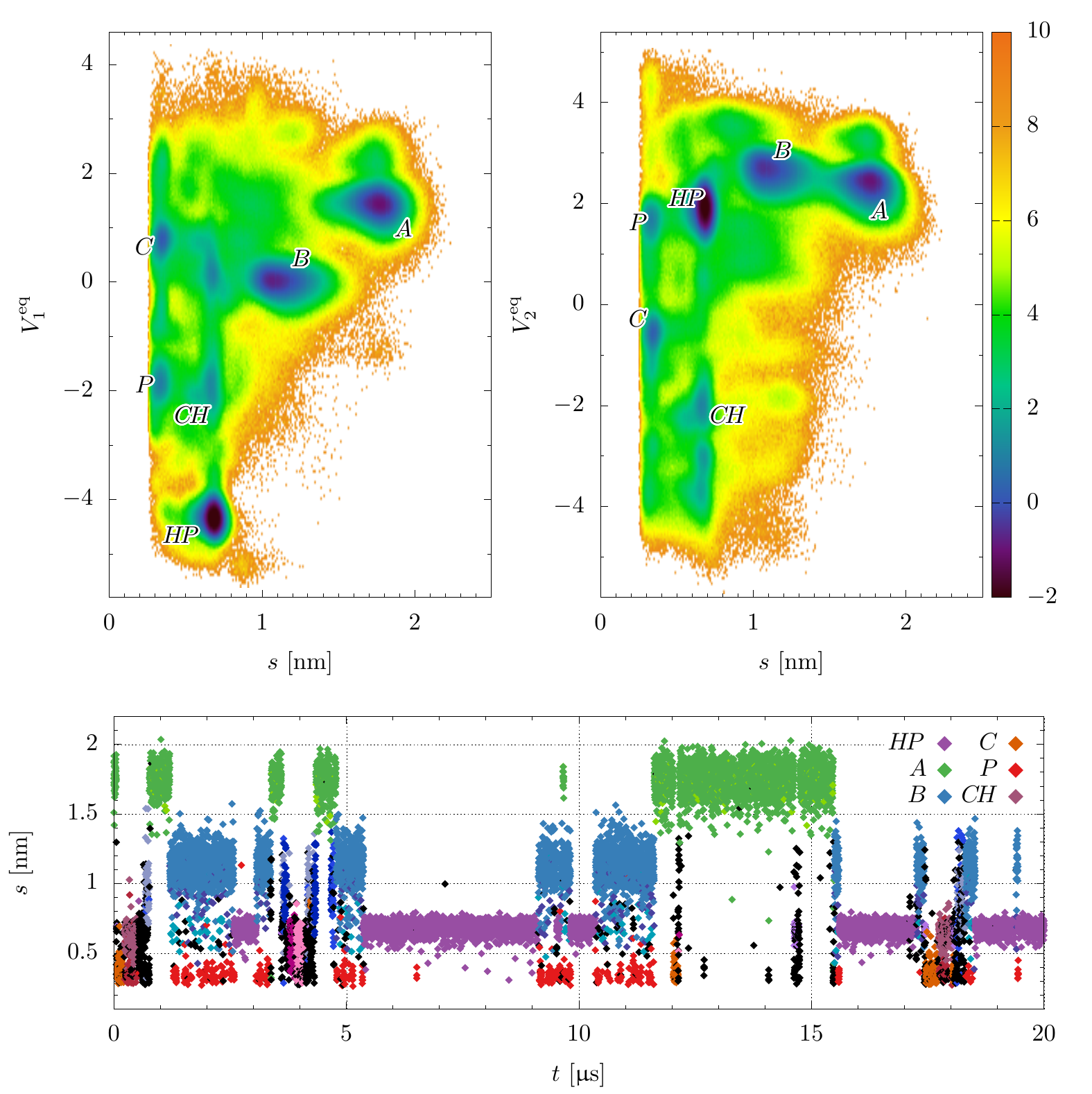}
	\caption{Performance of the distance $s$ between termini as reaction coordinate in case of the unbiased MD simulation of \ala. \textit{Top}: Free energy projected onto $s$ and one of the first two principle components. \textit{Bottom}: Time evolution of $s(t)$. The color coding is based on a density based state classification. There is a clear switching between the $P$ state (red) at $s \approx 0.35$ nm and the $B$ state (blue) at $s \approx 1.1$ nm. These transitions are hidden in the one-dimensional free energy profile $\Delta G(s)$.}
	\label{fig:eq_proj}
\end{figure}

\begin{figure}[h!]
	\centering
	\includegraphics{\dirfig/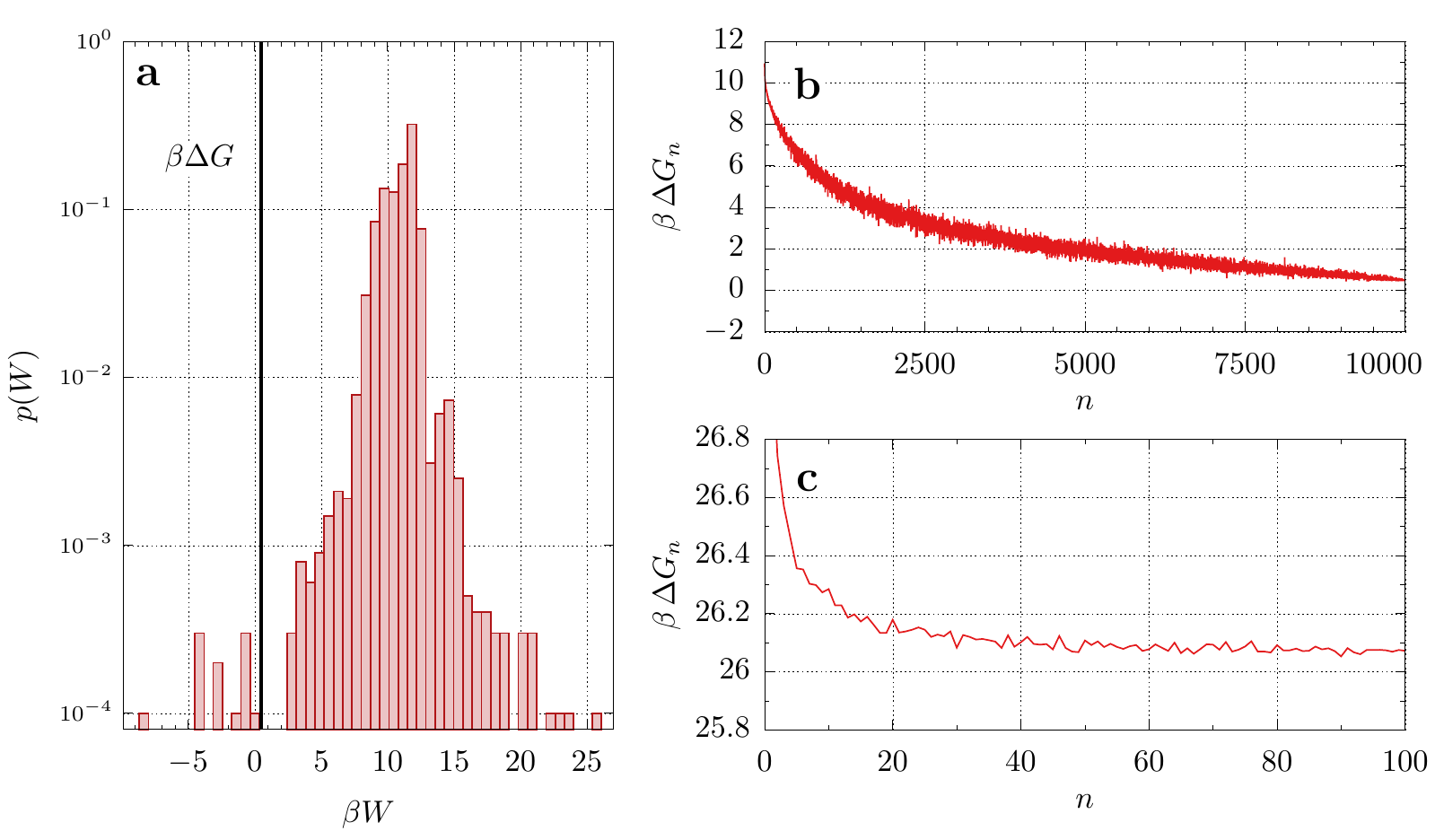}
	\caption{Convergence of the free energy estimate $\Delta G$ from Jarzynski's identity. (a) The work histogram (logscale) of the $B \rightarrow A$ transition at $s=1.77$ shows poor sampling for small values of $W$ even for a sample size of $K=10\, 000$ trajectories. (b) Block averaging of the free energy from the same data: Of the in total $K$ work values, $n$ are randomly chosen to estimate the free energy $\Delta G_{n,i}$, repeated $m=100 K/n$ times and averaged yielding $\Delta G_n = \frac{1}{m}\sum_{i}^m \Delta G_{n,i}$. The estimate $\Delta G_n$ is found to have not fully converged in the case for $B \rightarrow A$. (c) Same procedure repeated for the $A \rightarrow U$ transition at $s = 3$ nm. Here, only 100 trajectories are sufficient to converge to a plateau.}
	\label{fig:val_conv}
\end{figure}

\begin{figure}[h!]
	\centering
	\includegraphics{\dirfig/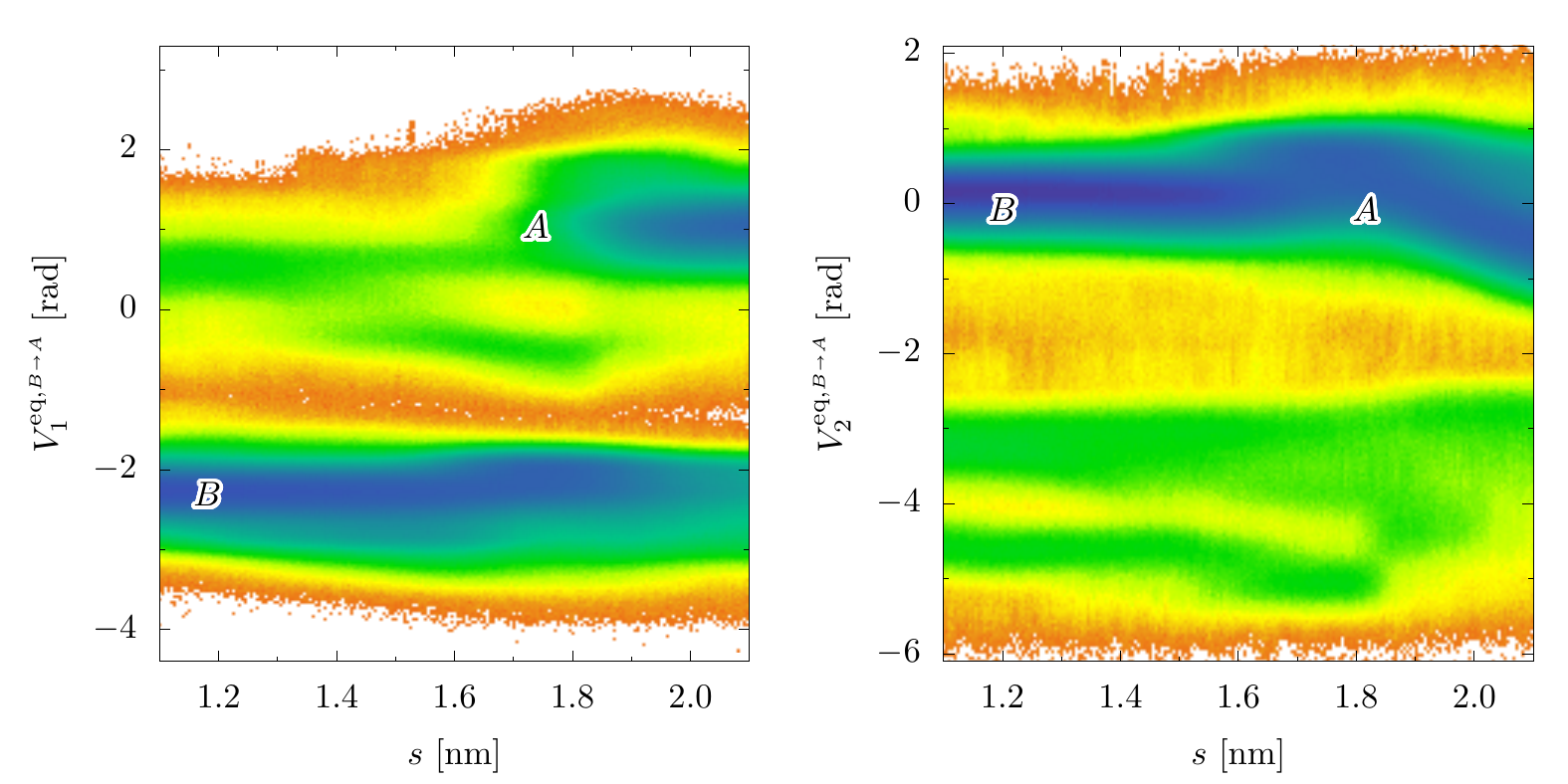}
	\caption{Energy landscape $\Delta \mathcal{G^{\text{neq}}}$ of the $B \rightarrow A$ transition of \ala, shown as a function of the pulling coordinate $s$ and one of the first principal components of the unbiased data $V^{\text{eq},B\shortrightarrow A}_{1,2}$. Evidently, a part of the trajectories does not make the transition to the $A$ state at all.}
	\label{fig:neq_pcs}
\end{figure}

\begin{figure}[h!]
	\centering
	\includegraphics{\dirfig/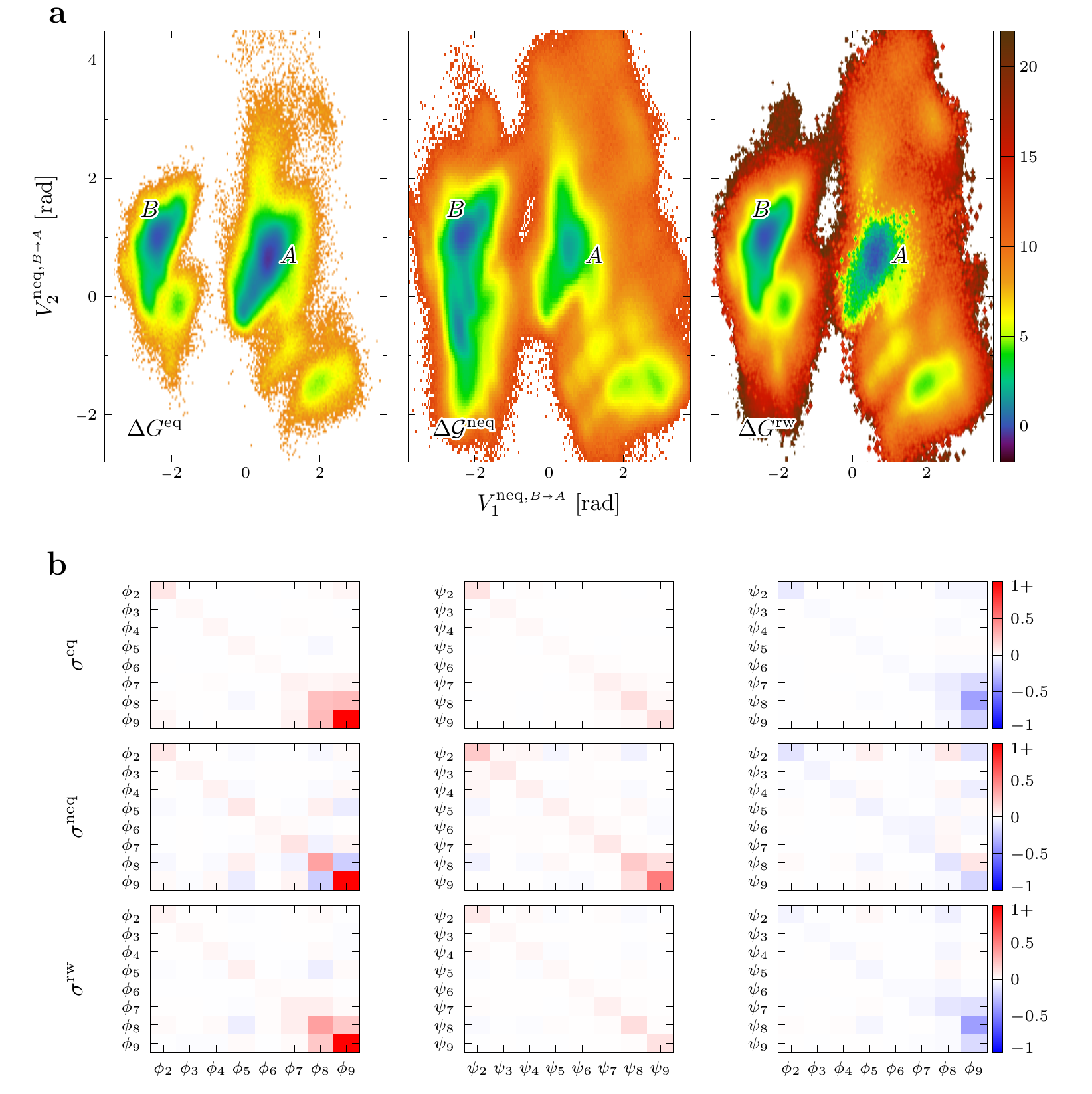}
	\caption{(a) Energy landscapes of the $B\rightarrow A$ transition of \ala, shown as a function the first two principal components $V_1^{\text{neq},B\shortrightarrow A}$ and $V_2^{\text{neq},B\shortrightarrow A}$ obtained from \dPCA~performed on the TMD data. (b) Comparison of the covariances of backbone dihedral angles of \ala~in the range $s = 1.1$ to $2.1$ nm reflecting the $B\rightarrow A$ transition, obtained from (top) unbiased MD simulation, (middle) TMD simulation and (bottom) reweighted TMD data. Note that the variance of the TMD data $\sigma^\text{neq}$ shows an anti-correlation of $\phi_8$ and $\phi_9$ caused by a different gap shift due to the constrained motion, correctly reweighted in $\sigma^\text{rw}$.}
	\label{fig:val_neqpcs}
\end{figure}

\begin{figure}[h!]
	\centering
	\includegraphics{\dirfig/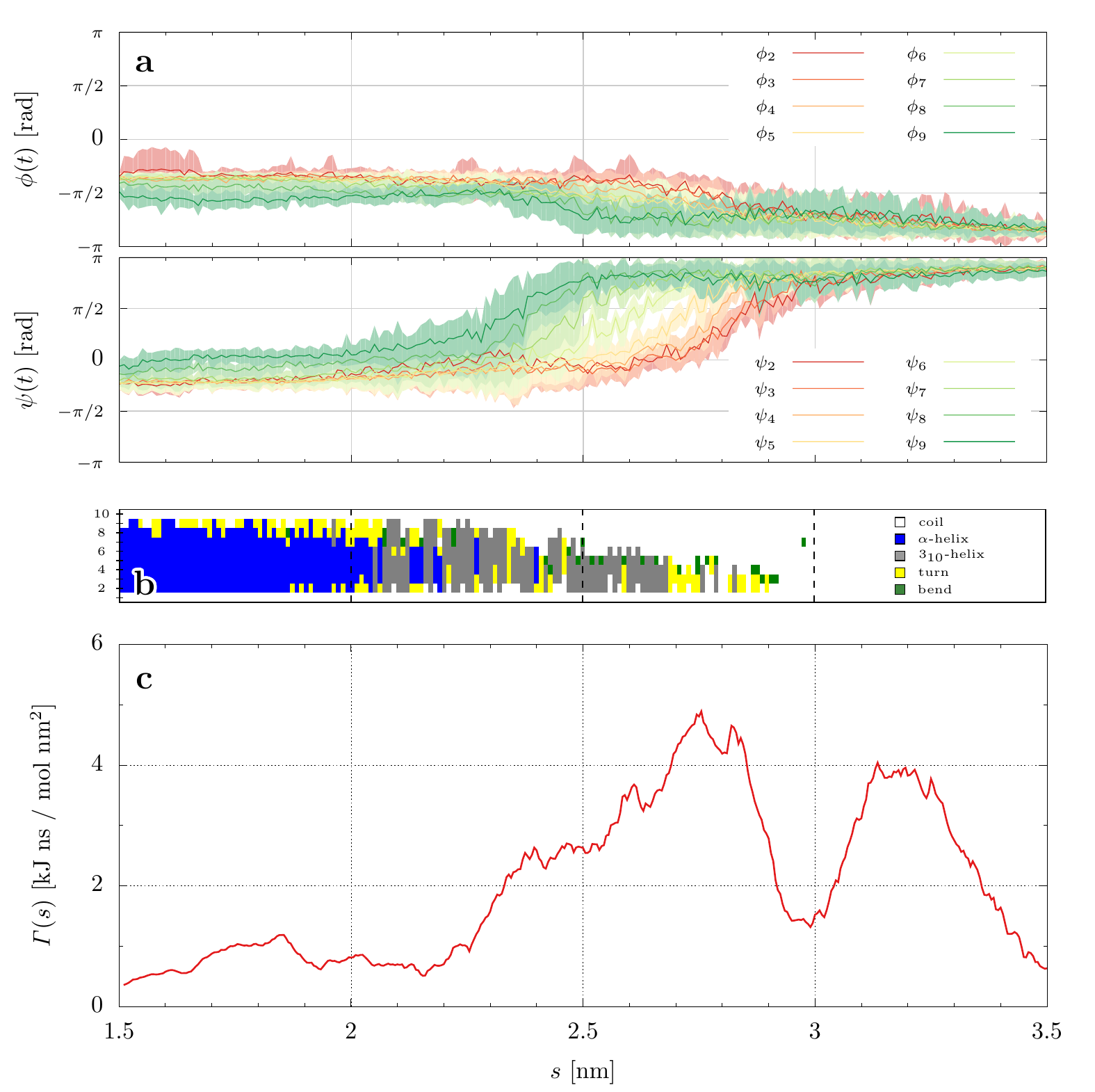}
	\caption{TMD simulations of $A\rightarrow U$ transition of \ala~as a function of the pulling coordinate $s$. Focusing on the bottom unfolding pathway shown in Fig.~3d of the main text, the peptide starts to unfold at the C-terminal and successively changes its dihedral angles until it is completely stretched. (a) Evolution of the backbone dihedral angles, where lines represent the mean values of the  dihedral angles, adjacent areas their variance. (b) Evolution of a (DSSP-based) secondary structure of a representative pulling trajectory, changing from $\alpha$-helix (blue) over $3_{10}$-helix (gray) to coil (white). (c) The corresponding friction profile $\itGamma(s)$ trends to increase during unfolding but becomes small after reaching full extension. $\itGamma(s)$ was smoothed by using a running average over 0.2 nm.}
	\label{fig:neq_dihe}
\end{figure}